\documentclass[twocolumn,english,aps,prl,reprint,nofootinbib,superscriptaddress]{revtex4-2}
\usepackage[T1]{fontenc}
\usepackage[latin9]{inputenc}
\usepackage{amsmath}
\usepackage{amssymb}
\usepackage{graphicx}

\usepackage{color}
\usepackage{float}
\usepackage{hyperref}
\usepackage{multirow}
\hypersetup{
    colorlinks,
    linkcolor={black},
    citecolor={black},
    urlcolor={blue}
}

\makeatletter
\@ifundefined{textcolor}{}
{%
 \definecolor{BLACK}{gray}{0}
 \definecolor{WHITE}{gray}{1}
 \definecolor{RED}{rgb}{1,0,0}
 \definecolor{GREEN}{rgb}{0,1,0}
 \definecolor{BLUE}{rgb}{0,0,1}
 \definecolor{CYAN}{cmyk}{1,0,0,0}
 \definecolor{MAGENTA}{cmyk}{0,1,0,0}
 \definecolor{YELLOW}{cmyk}{0,0,1,0}
 \definecolor{PURPLE}{rgb}{0.7,0,0.7}
 \definecolor{dgreen}{rgb}{0,0.6,0}
}
\usepackage{xspace}

\usepackage{pslatex}

\usepackage{babel}

\makeatother

\begin{document}

\title{Direct laser cooling of calcium monohydride molecules}

\author{S. F. V\'{a}zquez-Carson}
\affiliation{Department of Physics, Columbia University, New York, NY 10027-5255, USA}

\author{Q. Sun}
\affiliation{Department of Physics, Columbia University, New York, NY 10027-5255, USA}

\author{J. Dai}
\affiliation{Department of Physics, Columbia University, New York, NY 10027-5255, USA}

\author{D. Mitra}
\email{dm3710@columbia.edu}
\affiliation{Department of Physics, Columbia University, New York, NY 10027-5255, USA}

\author{T. Zelevinsky}
\email{tanya.zelevinsky@columbia.edu}
\affiliation{Department of Physics, Columbia University, New York, NY 10027-5255, USA}

\date{\today}

\begin{abstract}
\noindent 
We demonstrate optical cycling and sub-Doppler laser cooling of a cryogenic buffer-gas beam of calcium monohydride (CaH) molecules. We measure vibrational branching ratios for laser cooling transitions for both excited electronic states $A$ and $B$. We measure further that repeated photon scattering via the $A\leftarrow X$ transition is achievable at a rate of {$\sim1.6\times10^6$}~photons/s and demonstrate the interaction-time limited scattering of {$\sim200$} photons by repumping the largest vibrational decay channel. We also demonstrate the ability to sub-Doppler cool a molecular beam of CaH through the magnetically assisted Sisyphus effect. Using a standing wave of light, we lower the molecular beam's transverse temperature from {12.2(1.2)~mK to 5.7(1.1)~mK}. We compare these results to sub-Doppler forces modeled using optical Bloch equations and Monte Carlo simulations of the molecular beam trajectories. This work establishes a clear pathway for creating a magneto-optical trap (MOT) of CaH molecules. Such a MOT could serve as a starting point for production of ultracold hydrogen gas via dissociation of a trapped CaH cloud.
\end{abstract}
\maketitle

\section{Introduction}\label{sec:intro}
The development of robust techniques for laser cooling of atoms \cite{Raab_1987_mot,Phillips_Nobel_1998} has led to major advancements in the fields of quantum simulation, quantum computation, and frequency metrology, and has enabled precise tests of fundamental physics \cite{Bloch_ManyBodyGas_2008,Morgado_2021_Rydberg_qubits,YeBothwellNature22_mmRedshift,Safronova_NewPhysicsAtoms_2018}. Cooling and trapping of molecules represents the next level in experimental complexity because of the additional internal degrees of freedom and lack of perfect two-level structure that can be used for sustained photon cycling \cite{Tarbutt_LaserCooling_2018}. In exchange for this increase in complexity, molecules provide enhanced sensitivity for fundamental precision measurements \cite{ACMENature18_ACMEIIeEDM,YeCairncrossNatRevPhys19_TSymmAtomsMolecules,Augenbraun_PolyatomicEDMSearch_2020}, longer coherence times for quantum information \cite{Park_NaKCoherence_2017, CornishBlackmoreQST19_RotationalCoherence_CaF_RbCs, Sawant_UltracoldMolQubit_2020}, and tunable long-range interactions for quantum simulators \cite{Hazzard_DipolarMolManyBody_2014,CornishBlackmoreQST19_RotationalCoherence_CaF_RbCs}. The technique of buffer gas cooling \cite{Hutzler_CR2012_BufferGasBeams,DeMilleBarryPCCP11_CryogenicMolecularBeams} has enabled direct laser cooling of molecules, including several diatomic \cite{Shuman_LaserCoolingDiatomic_2010, Truppe_DopplerLimitMolCooled_2017, Anderegg_RFMotCaF_2017, Collopy_YOMot_2018,ZelevinskyMcNallyNJP20_BaH1DCooling}, triatomic \cite{Kozyryev_SisyphusSrOH_2017, Augenbraun_PolyatomicEDMSearch_2020, Vilas_arXiv2021_3D_MOT_CaOH}, and symmetric top \cite{Mitra_CaOCH3Sisphus_2020} species. We add the alkaline-earth monohydride CaH to this growing list.

A cold and trapped cloud of hydrogen atoms promises to be an ideal system for testing quantum electrodynamics (QED) and precise measurements of fundamental constants
\cite{CODATA2018RMP21,Biraben_2009_spectroscopy_hydrogen}. More than two decades ago, a BEC of atomic hydrogen was prepared in a magnetic trap \cite{Fried_HydrogenBEC_1998}. The measurement of the $1S-2S$ transition has been performed in a magnetic trap of hydrogen \cite{Cesar_1996_hydrogen_magnetic_trap} and antihydrogen \cite{Ahmadi_Nature2017_antihydrogen_1S2S_trapped} and also in a beam \cite{Parthey_PRL2011_Hydrogen_1s2s_beam}. More recently, experiments measured the $1S-3S$ \cite{Grinin_Science2020_Hydrogen_1S3S_beam} and the $2S_{1/2}-8D_{5/2}$ \cite{Brandt_PRL2022_hydrogen_1S5D_beam} transitions of hydrogen with unprecedented precision. Furthermore, magnetic slowing of paramagnetic hydrogen has been proposed \cite{ Raizen_AtomicMotion_2009}. While extremely successful, these experiments are limited by
motional effects such as collisions. The ability to perform these measurements with a dilute ultracold sample of hydrogen tightly trapped in an optical potential could significantly improve the precision. 

One promising pathway that was proposed is via the fragmentation of hydride molecules \cite{Lane_UltrcoldHydrogen_2015} and ions \cite{Jones_2022_BaH_ions}. Diatomic hydride radicals can be efficiently cooled using direct laser cooling techniques. Additionally, if the fragmentation process is not exothermic (i.e., the binding energy is carried away by an emitted photon), the resulting hydrogen atoms can populate a Boltzmann distribution at a lower temperature than the parent molecules. This presents barium monohydride as an ideal candidate where the large mass difference between the barium atom and the hydrogen atom could result in an ultracold cloud of hydrogen atoms after fragmentation. Although BaH was successfully laser cooled \cite{ZelevinskyMcNallyNJP20_BaH1DCooling}, its low recoil momentum and a relatively weak radiation pressure force make it challenging to load BaH in a magneto-optical trap. 

In this work, we explore another hydride candidate for laser cooling and trapping, calcium monohydride. Due to its diode-laser accessible transitions and short excited state lifetimes, CaH is a promising candidate for optical cycling. In Sec. \ref{sec:setup}, we describe the electronic structure and useful transitions for this molecule. We also characterize our cryogenic buffer gas beam source. In Sec. \ref{sec:VBR}, we summarize the measurement of the vibrational branching ratios for this molecule and establish a photon budget for laser cooling. In Sec. \ref{sec:scattrate}, we present the measurement of the photon scattering rate and show that we can achieve rates over $10^6$~photons$/$s. In Sec. \ref{sec:sisyphus}, we demonstrate our ability to sub-Doppler cool a beam of CaH by {$\sim2\times$} in one dimension while only scattering {$\sim$ 140}~photons. Finally, in Sec. \ref{Sec:Conclusion} we conclude that these results establish CaH as a promising candidate for laser cooling and trapping.     

\begin{figure*}[ht!]
\includegraphics[scale = 0.75]{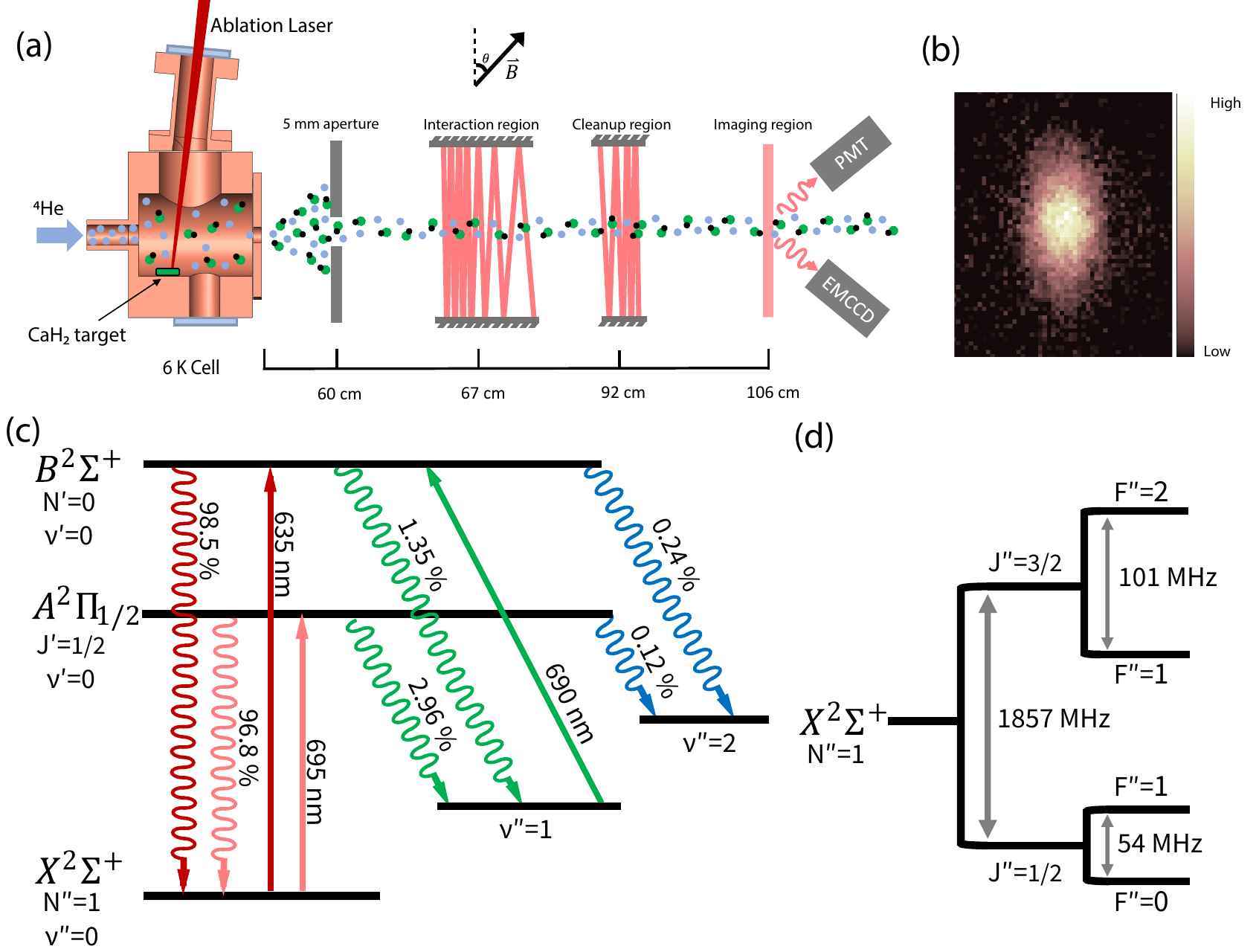}
\caption{Experimental setup and molecular structure. (a) Experiment diagram (not to scale). CaH molecules are produced via ablation of a solid target of CaH$_{2}$. The ejected molecules thermalize to the $\sim6$~K $^4$He buffer gas and are extracted through a 3~mm diameter aperture. An additional 5~mm aperture is placed just before the molecular beam enters the interaction region to limit the transverse velocity distribution. The cooling chamber consists of 12~cm of optical access followed by a cleanup region and a detection region. (b) A sample image of the molecular cloud detected using the EMCCD camera. (c) The main cooling line used in this work is the $A^2 \Pi_{1/2} (\nu'=0,\ J'=1/2)$  $\leftarrow$ $X^2 \Sigma^{+} (\nu''=0,\ N''=1)$ transition at 695 nm. We detect the molecular beam using the $B^2 \Sigma^{+} (\nu'=0,\ N'=0)$ $\leftarrow$ $X^2 \Sigma^{+} (\nu''=0,\ N''=1)$ transition at 635 nm. We employ laser light at 690 nm to repump molecules that decay to the $\nu''=1$ vibrational state in $X$ by addressing the $B^2 \Sigma^{+} (\nu'=0,\ N'=0)$ $\leftarrow$ $X^2 \Sigma^{+} (\nu''=1,\ N''=1)$ transition. By closing this leak, we create a quasi-closed transition capable of cycling $\sim 200$ photons sufficient to exert a measurable Sisyphus force. (d) The 1.86 GHz spin-rotation splitting and the 54~MHz~(101~MHz) hyperfine structure of the $J=1/2~(J=3/2)$ states in the ground $X$ manifold. The details of the laser setup can be found in Appendix \ref{sec:lasers}.}
\label{fig:Mol_Structure}
\end{figure*}

\section{Experimental setup}\label{sec:setup}
The experiment consists of a cryogenic buffer gas beam source
operating at $\sim6$~K \cite{Hutzler_CR2012_BufferGasBeams,DeMilleBarryPCCP11_CryogenicMolecularBeams,ZelevinskyMcNallyNJP20_BaH1DCooling}. We employ $^4$He as a buffer gas that is flown into the cell at $\sim$ 6~sccm (standard cubic centimeter per minute) flow rate via a capillary on the back of the cell (Fig. \ref{fig:Mol_Structure}(a)). The target is composed of pieces of CaH$_2$ (Sigma-Aldrich, 95$\%$ purity) held on a copper stub using epoxy. To ablate the target, we use the fundamental output of an Nd:YAG pulsed laser operating at 1064~nm and at a 2~Hz repetition rate. We run the ablation laser at a maximum pulse energy of 30~mJ and focus the beam to a 1.5~mm diameter. We observe the highest molecular yield when the ablation energy is deposited over a large target surface area. The CaH radicals produced due to ablation subsequently thermalize their internal rotational and vibrational degrees of freedom via collisions with the buffer gas. These molecules are then hydrodynamically entrained in the buffer gas flow out of the cell. The molecules leaving the cell are predominantly in the lowest two rotational states ($N''=0$ and $1$) and the ground vibrational state ($\nu''=0$).

After leaving the cryostat, the molecules enter a high vacuum chamber equipped with a beam aperture of 5~mm diameter to filter out the $1/e^2$ transverse velocity range to {$\sim\pm$3~m/s} (Fig. \ref{fig:Mol_Structure}(a)). We keep the aperture in place for all data shown in this work. Subsequently, the molecules enter an interaction region with rectangular, antireflection coated windows enabling a 12~cm long interaction length. Next, the molecules enter a ``cleanup'' region, where population accumulated in the $X(\nu''=1)$ state is pumped back to the $X(\nu''=0)$ state via the $B(\nu'=0)$ state, and are then detected in the imaging region by scattering photons on the $B(\nu'=0)\leftarrow$ $X(\nu''=0)$ transition. The scattered photons are simultaneously collected on a photon counting photo-multiplier tube (PMT) and an electron-multiplying charge-coupled device (EMCCD) camera. {An example of the average camera images collected is shown in Fig. \ref{fig:Mol_Structure}(b).}

The relevant energy level structure for CaH is depicted in Fig. \ref{fig:Mol_Structure}(c). We start in the ground electronic manifold ($X^2\Sigma^+$) and excite to the two lowest excited electronic states ($A^2\Pi_{1/2}$ and $B^2\Sigma^+$). A rotationally closed optical cycling transition can be guaranteed by selection rules if we address the $N''=1,\ J''=1/2, 3/2$ ground states to the opposite parity $J'=1/2$ excited state in the $A$ manifold or the $N'=0$ state in the $B$ manifold \cite{YeStuhlPRL08_PolarMoleculeMOT}. The ground $X(N=1)$ state is split into two components separated by 1.86~GHz due to the spin-rotation interaction (Fig.\ref{fig:Mol_Structure}(d)). Each sublevel is further split into two hyperfine sublevels ($J=1/2,\ F=0,1$ and $J=3/2,\ F=1,2$) separated by 54~MHz and 101~MHz respectively. Each hyperfine sublevel is composed of $2F+1$ $m_F$ states that remain unresolved for the purpose of this study. The primary vibrational decay from both $A$ and $B$ excited states is to the $\nu''=1$ state ($\sim1$-$3\%$) and subsequently to the $\nu''=2$ state ($< 0.3\%$). For this work, we only repump the population out of the $\nu''=1$ state. The details of the laser setup can be found in Appendix \ref{sec:lasers}. 

We estimate the longitudinal velocity of our molecular beam as follows. Every ablation pulse also produces a sizeable number of Ca atoms that simultaneously get buffer gas cooled and extracted from the cell alongside the CaH molecules. We measure the longitudinal velocity profile of these Ca atoms by addressing the $^1S_0 \rightarrow ^1P_1$ transition in calcium at 423~nm in a velocity sensitive configuration. The high density of calcium in the beam allows us to measure the Doppler-shifted atomic resonance with a high signal-to-noise ratio. We find that the longitudinal velocity is peaked at {$\sim250$~m/s} with a full width at half maximum of {$\sim200$~m/s}. Since the masses of CaH and Ca are nearly identical and they experience identical buffer-gas cooling, we assign the same longitudinal velocity profile to both species. We also estimate the molecular beam flux in our system via the total camera counts and the estimated collection efficiency of the imaging system. We obtain a typical beam flux of {$\sim1\times10^{10}$} molecules$/$steradian$/$pulse.

\section{Vibrational Branching Ratio Measurement}
\label{sec:VBR}

Although the cycling transition in CaH is rotationally closed, no selection rules prevent vibrational decay. The probability of decay from the excited electronic state to vibrationally excited $X$ states is quantified by the vibrational branching ratio (VBR). To reduce the number of repumping lasers required to scatter $\sim 10^5$ photons, it is essential that the off-diagonal VBRs are highly suppressed \cite{DiRosaEPJD04_LaserCoolingMolecules,YeStuhlPRL08_PolarMoleculeMOT}. The directly laser cooled diatomic molecules to date, including  CaF \cite{Truppe_DopplerLimitMolCooled_2017,Anderegg_RFMotCaF_2017}, SrF \cite{Barry_2014_SrFMOT}, YbF \cite{Lim_CooledYbF_2018} and YO \cite{Collopy_YOMot_2018}, all possess highly diagonal transitions. The Franck-Condon factor (FCF) is defined as the square of the wavefunction overlap of two different vibrational states. VBRs (denoted as $q$) can be calculated from FCFs (denoted as $f$) using  
\begin{equation}
    \label{eq:fcf2vbr}
    q_{\nu' \nu''} = \frac{f_{\nu'\nu''} \times \omega^3_{\nu'\nu''}}{\sum_{n=0}^{\infty} f_{\nu'n} \times \omega^3_{\nu'n}}
\end{equation}
\noindent where $\omega_{\nu'\nu''}$ is the positive energy difference between the states $\nu'$ and $\nu''$. Alkaline-earth monohydrides have been extensively studied and FCFs have been calculated or measured for BeH, MgH, SrH and BaH \cite{Gao_Summary_2014, Cheng_MgH_FCF_2015, Tarallo_BaHFCF_2016, Ramanaiah_1982_CaH_FCF}. Calculated values for CaH \cite{Gao_Summary_2014,Pathak_1966_CaH_FCF,Ramanaiah_1982_CaH_FCF} are summarized in Table \ref{tab:VBRs}. In this section we report our measurement of VBRs for the $A(\nu'=0)$ and $B(\nu'=0)$ states of CaH, denoted by $q_{0,\nu''}$ where $\nu''$ is 0, 1 and 2.

We perform the VBR measurement with our molecular beam using a process similar to the one described in Ref. \cite{Hendricks_2014_VBR_BF}. A pump laser beam intersects the CaH beam orthogonally in the imaging region, and resonantly excites the molecules from $X^2 \Sigma^{+} (\nu''=0)$ ground states to the $A^2 \Pi_{1/2} (\nu'=0)$ or $B^2 \Sigma^+ (\nu'=0)$ excited states. Once excited, two PMTs in photon-counting mode with different dichroic filters are used to collect the photons simultaneously emitted from the various decay pathways of the excited state. The narrow bandpass dichroic filters are strategically chosen to isolate photons with a frequency resonant with vibrational decay to a single excited vibrational state $(\lambda _{0\nu''})$ while simultaneously detecting the molecules that return to the ground state $(\lambda _{00})$\footnote{A complete list of filters used in this experiment and their measured transmission efficiencies at certain wavelengths can be found in Sec. \ref{sec:filters}.}. 

\begin{figure}
\includegraphics[scale=1]{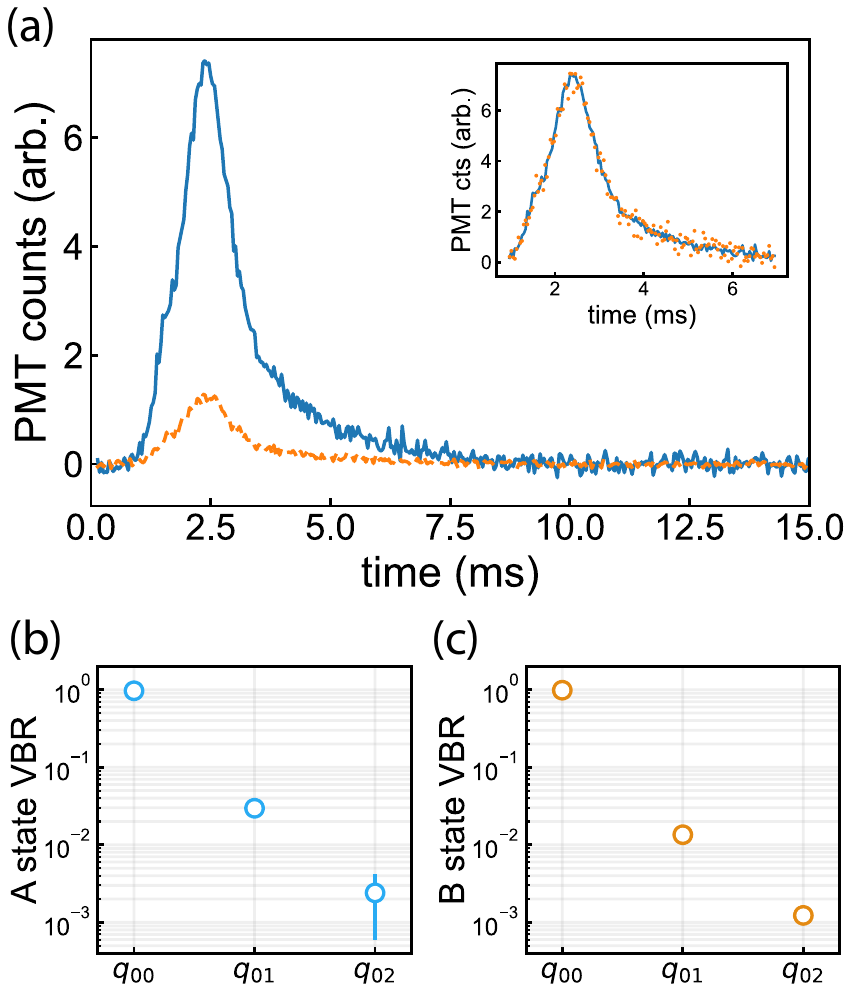}
\caption{Measurement of the VBRs for CaH molecules. (a) An example of PMT traces used to calculate the VBRs. These two time traces correspond to background-subtracted fluorescence from the decay to $X(\nu''=0)$ for PMTs $P_1$ (orange, {dashed}) and $P_2$ (blue, {solid}) while the $B^2\Sigma^+(\nu'=0)\leftarrow X^2\Sigma^+(\nu''=0)$ transition is excited. The inset denotes how the ratio of integrated signals, $R_0$, is computed. We perform a one-parameter fit of the dashed trace to the solid trace. The result of the fit produces the orange points that can be seen to overlap temporally with the blue trace. (b,c) The resulting VBRs from the measured ratio $R_{\nu''}$, obtained by evaluating Eq. (\ref{eq:A}), for the $A$ and $B$ states. Each point represents an average of at least 200 shots with background subtraction, while the higher vibrational decays require  $\sim700$ shots for an appreciable signal-to-noise ratio due to the low probability of decaying to these excited states. Error bars are statistical standard errors.
}
\label{fig:fcf}
\end{figure}

\begin{table*}[htbp]
\centering
\begin{tabular}{|c|c|c|c|c|c|c|}
\hline
Transition     &  \begin{tabular}[c]{@{}c@{}} Lifetime\\$\tau$ (ns)\end{tabular}  & \begin{tabular}[c]{@{}c@{}} Vibrational Quanta\\ ($\nu''$)\end{tabular} & \begin{tabular}[c]{@{}c@{}}Transition wavelength\\ (nm)\end{tabular} & \begin{tabular}[c]{@{}c@{}}FCF Theory \\ ($f_{0\nu''}$)\end{tabular}                   & \begin{tabular}[c]{@{}c@{}}FCF measured \\ ($f_{0\nu''}$)\end{tabular} & \begin{tabular}[c]{@{}c@{}}VBR measured\\ ($q_{0\nu''}$)\end{tabular}\\ \hline
\multirow{4}{*}{$A \rightarrow X$} & \multirow{4}{*}{33(3)} & 0 & 695.13 & 0.953 & 0.9572(43)  &  0.9680(29)   \\ \cline{3-7} 
                    & & 1 & 761.87 &  0.0439  & 0.0386(32)   & 0.0296(24)    \\ \cline{3-7} 
                    & & 2& 840.07 &2.74$\times10^{-3}$&4.2(3.2)$\times10^{-3}$ & 2.4(1.8)$\times10^{-3}$ \\ \cline{3-7}
                    & & 3 & 932.80 &  2.3$\times10^{-4}$  & -  &   -   \\\hline
\multirow{4}{*}{$B \rightarrow X$} & \multirow{4}{*}{58(2)} & 0  & 635.12   &  0.9856  & 0.9807(13)  &  0.9853(11)\\ \cline{3-7} 
                    & & 1  &  690.37    &  0.0132  & 0.0173(13) & 0.0135(11)    \\ \cline{3-7} 
                    & & 2  & 753.97  & 1.1$\times10^{-3}$&2.0(0.3)$\times10^{-3}$& 1.2(0.2)$\times10^{-3}$\\ \cline{3-7}
                    & & 3  &  827.84   &  1$\times10^{-4}$  & - & -     \\ \hline
\end{tabular}
\caption{FCFs and VBRs for the measured transitions of CaH. Measured excited state radiative lifetime for the $A$ state was obtained from Ref. \cite{Liu_2009_CaH_A_lifetime} and for the $B$ state from Ref. \cite{Berg_1996_CaH_B_Lifetime}. The excited state vibrational quantum is always $\nu'=0$. The $A(\nu'=0) \leftarrow X(\nu''=0)$ excitation wavelength at 695.13~nm, the $B(\nu'=0) \leftarrow X(\nu''=0)$ excitation wavelength at 635.12~nm, and the $B(\nu'=0) \leftarrow X(\nu''=1)$ excitation wavelength at 690.37~nm were determined experimentally. The other transition wavelengths are derived using measured vibrational energies given in Ref. \cite{Shayesteh_2013_CaH_fourier_spectra}. The calculated FCFs for the $A \rightarrow X$ decay are obtained from \cite{Gao_Summary_2014} while for the $B \rightarrow X$ decay they are obtained from \cite{Ramanaiah_1982_CaH_FCF}. Error bars for the measured FCFs and VBRs are statistical standard errors.
}
\label{tab:VBRs}
\end{table*}

We first compare the time traces of two PMTs when their filters allow transmission at the same $\lambda_{00}$ frequency. The ratio of integrated signals, $R_0$, can be expressed with systematic parameters and VBRs as

\begin{equation}
     \label{eq:R1}
     R_0 = \frac{N q_{00} \Omega_{P_2} T_{F_2,\lambda_{00}} Q_{P_2,\lambda_{00}}}{N q_{00} \Omega_{P_1} T_{F_1,\lambda_{00}} Q_{P_1,\lambda_{00}} },
\end{equation}

\noindent where the subscripts $P_1/P_2$ stand for two PMTs used in this experiment\footnote{PMTs used: Hamamatsu R13456 and SensTech P30PC-01.}, subscripts $F_1/F_2$ stand for the two bandpass filters used, $N$ is the number of scattering events, $q_{00}$ is the diagonal VBR, $\Omega_P$ is the geometrical collection efficiency for a given PMT, $T_{F,\lambda}$ is the transmission efficiency for a given bandpass filter at a wavelength $\lambda$, and $Q_{P,\lambda}$ is the quantum efficiency for a given PMT at a wavelength $\lambda$. 

Next, we replace filter $F_2$ with another filter $F_3$ which blocks transmission at $\lambda_{00}$ and allows transmission at $\lambda_{0\nu''}$, where $\nu''$ is 1 or 2. The ratio of integrated signals, $R_{\nu''}$, can then be written
as
\begin{equation}
     \label{eq:R2}
     R_{\nu''} = \frac{N' q_{0\nu''} \Omega_{P_2} T_{F_3,\lambda_{0\nu''}} Q_{P_2,\lambda_{0\nu''}}}{N' q_{00} \Omega_{P_1} T_{F_1,\lambda_{00}} Q_{P_1,\lambda_{00}} }.
\end{equation}

An example of the measured signals is shown in Fig. \ref{fig:fcf}(a). For each measurement we simultaneously collect the time traces from the two
PMTs. In order to obtain the ratio $R_{\nu''}$, we perform a one-parameter least square fit of all points in one time trace to the other (Fig. \ref{fig:fcf}(a) inset). Since the PMTs are stationary throughout the experiment, any variation in $\Omega$ is negligible. By measuring the transmission efficiency of the filters $F_2 / F_3$ at $\lambda_{00} / \lambda_{0\nu''}$, as well as the quantum efficiency of $P_2$ at $\lambda_{00} / \lambda_{0\nu''}$, and combining Eqs. (\ref{eq:R1}) and (\ref{eq:R2}), we estimate the ratio of VBRs as
\begin{equation}
     \label{eq:A}
     \frac{q_{0\nu''}}{q_{00}} = \frac{R_{\nu''} Q_{{P_2}, \lambda_{00}} T_{F_2, \lambda_{00}}}{R_0 Q_{{P_2}, \lambda_{0\nu''}} T_{F_3, \lambda_{0\nu''}}}.
\end{equation}

We calculate the individual VBRs by assuming that the sum is $\sum_{\nu''=0}^2\ q_{0\nu''}=1$. This is a reasonable approximation since the calculated value of $f_{03}$ is smaller than the statistical uncertainty in the measured FCFs for both $A$ and $B$ states (Table \ref{tab:VBRs}). The resulting VBRs are plotted in Figs. \ref{fig:fcf}(b,c). The measured FCFs were calculated using the inverted form of Eq. (\ref{eq:fcf2vbr}).
\section{Scattering Rate Measurement}
\label{sec:scattrate}

Efficient cooling and slowing of molecules require rapid scattering of photons while simultaneously minimizing the loss to unaddressed vibrationally excited states. From the measured VBRs for the primary decay pathways for CaH as described in Sec. \ref{sec:VBR}, we obtain the average number of photons per molecule, $\langle N_{\mathrm{ph}}\rangle$, that we expect to scatter while addressing $N_v$ vibrational channels before only $1/e$ of the ground state population remains available for optical cycling as 
\begin{equation}
     \label{eq:N_Phot}
     \langle N_{\mathrm{ph}} \rangle \simeq \frac{1}{1-\sum_{\nu''=0}^{N_v}q_{0\nu''}}.
\end{equation}

Thus we expect to scatter {31(3)} photons for $A\leftarrow X(\nu''=0)$ and {68(5)} photons for $B\leftarrow X(\nu''=0)$ before losing $63\%$ of molecules to the $X(\nu''=1)$ state. Next, if the $X(\nu''=1)$ state is repumped, this photon number increases to around {400} for the $A$ state and {800} for the $B$ state cycling schemes. In order to slow a CaH molecule travelling at 250~m/s to within the capture velocity of a MOT \cite{Anderegg_RFMotCaF_2017, ZelevinskyMcNallyNJP20_BaH1DCooling}, we would need to scatter $\sim2\times10^{4}$ photons. Although the loss to excited vibrational modes can be minimized by using repumping lasers for higher vibrational states, it is essential to scatter photons at a high rate so that the slowing distance can be minimized. The maximum scattering rate for a multilevel system with $n_g$ ground states and $n_e$ excited states is given by \cite{Norrgard_2016_RFMOT}
\begin{equation}
     \label{eq:Max_Scat}
     R_{\mathrm{sc},\mathrm{max}}=\Gamma_{\mathrm{eff}} = \frac{1}{\tau} \frac{n_e}{n_e + n_g}
\end{equation}

\noindent where
$\tau$ is the excited state lifetime given in Table \ref{tab:VBRs}. The rotationally closed transition employed here is $N'=0\leftarrow N''=1$, i.e., $n_e = 4$ and $n_g = 12$ (see Fig. \ref{fig:Mol_Structure}(d)). Here we assume that the repumping lasers couple to different excited states. We obtain the maximum scattering rate {$\sim7.6\times10^6$~s$^{-1}$} for the $A$ state and {$\sim4.3\times10^6$~s$^{-1}$} for the $B$ state. In practice, however, it is difficult to achieve these maximum values and most experiments with diatomic, triatomic, and polyatomic molecules to date achieve scattering rates up to $\sim2\times10^6$~s$^{-1}$.

\begin{figure}[htbp]
\includegraphics[scale=0.75]{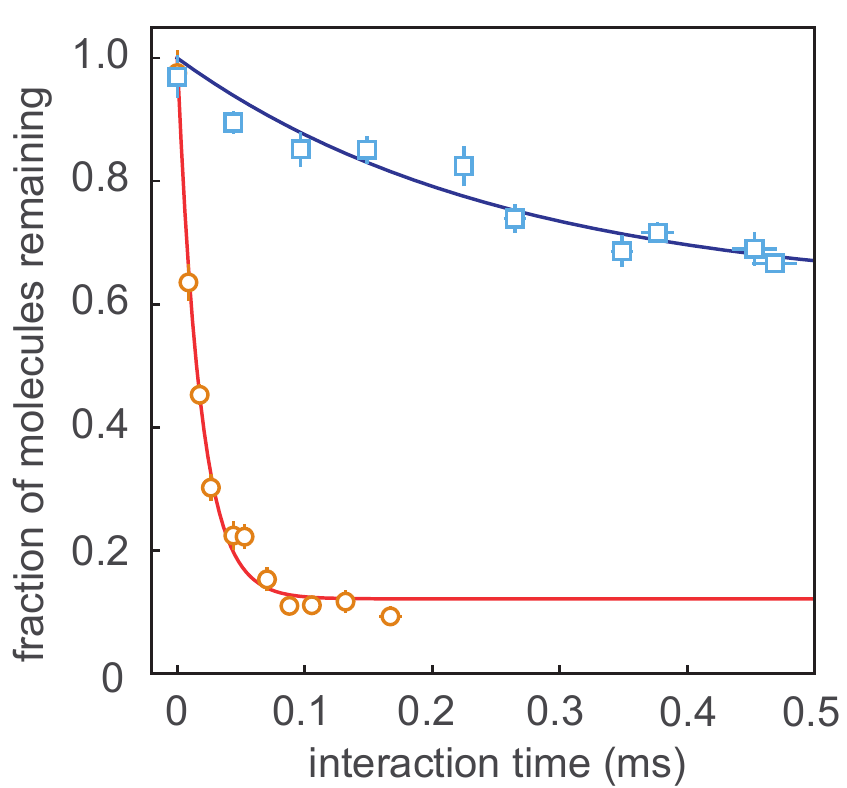}
\caption{Scattering rate measurement. The fraction of molecules remaining in the $X(\nu''=0)$ state when cycling on the $A (\nu'=0)\leftarrow X(\nu''=0)$ transition ({orange circles}) and the fraction remaining in $X(\nu''=0)+X(\nu''=1)$ states when cycling simultaneously on the $A(\nu'=0)\leftarrow X(\nu''=0)$ and the $B(\nu'=0)\leftarrow X(\nu''=1)$ transitions ({blue squares}) are measured as a function of the interaction time. Lines are fits to exponential decay curves with finite offsets. The offset is due to detected molecules that are only weakly addressed in the interaction region.}
\label{fig:scatt_rate}
\end{figure}

In order to measure the maximum scattering rate achievable in our setup, we measure the fraction of molecules that are pumped to dark vibrationally excited states as a function of interaction time. First, we apply only the $A\leftarrow X(\nu''=0)$ linearly polarized, resonant light ({$\sim80$~mW} per spin-rotation component) in the interaction region in a multi-pass configuration (Fig.\ref{fig:Mol_Structure}(a)). Each pass of the laser beam is spatially resolved so that the effective interaction length can be varied and quantified by counting the number of passes. We measure the population remaining in $X(\nu''=0)$ and convert the interaction length to time by measuring the laser beam waist ($1/e^2$ radius of {$0.55$~mm} in the direction parallel to the molecular beam and {0.84}~mm in the orthogonal direction) and approximating that each molecule travels with a {250~m/s} longitudinal velocity. We also apply a 3~G magnetic field in the interaction region to destabilize the dark magnetic sublevels that become populated during optical cycling. Magnetic field strength and laser polarization angle with respect to the magnetic field are scanned to maximize the scattering rate. The angle between the magnetic field and the polarization of the laser that addresses $|X,\ J=3/2,\ F=2\rangle$
was ultimately chosen to be $\sim13^{\circ}$.

As the molecules propagate through the interaction region and scatter photons, some of the excited state molecules decay to unaddressed higher vibrational states at a rate given by the sum of addressed state VBRs as
\begin{equation}
    f_{\mathrm{rem}}(t) = \frac{N_{\mathrm{mol}}(t)}{N_{\mathrm{mol}}(t=0)} = \left(\sum_{\nu''=0}^{v_a}q_{0\nu''}\right)^{N_p(t)},
\end{equation}

\noindent where $f_{\mathrm{rem}}$ is the fraction of molecules that remain in all the addressed states combined. The number of scattered photons is $N_p(t) = R_{\mathrm{sc}}t$, and $v_a$ is the highest addressed vibrational level. The experimental data is shown in Fig. \ref{fig:scatt_rate} (orange circles). We fit the decay in $f_{\mathrm{rem}}$ to an exponential decay with a finite offset. We note that in the limit of infinite interaction time, $f_{\mathrm{rem}}\rightarrow0$. However, in our setup we have a small fraction of the molecules that only weakly interact with the laser beam but are still detected in the imaging region. These molecules are accounted for by adding a constant offset to $f_{\mathrm{rem}}$. From the exponential decay constant $\tau_d$, we can obtain the scattering rate  
\begin{equation}
     \label{eq:ScatFit}
     R_{\mathrm{sc}} \simeq \frac{1}{\tau_d\left(1-\sum_{\nu''=0}^{v_a}q_{0\nu''}\right)}.
\end{equation}
Using our measured values of the VBRs and $\tau_d$ from the orange curve of Fig. \ref{fig:scatt_rate},
we estimate a maximum scattering rate of {$1.67(15)\times10^6$~s$^{-1}$}. Here we make a simplifying assumption that the laser intensity sufficiently exceeds the saturation intensity
such that the local variation in intensity due to the Gaussian laser beam profile does not affect our estimate.   

Next, we measure $R_{\mathrm{sc}}$
after adding {$\sim110$~mW} of repumping light addressing the $B(\nu'=0)\leftarrow X(\nu''=1)$ transition, co-propagating with the main cycling light. In this case, we also add $\sim40$~mW of the same repumping light to the cleanup region. Within this multi-pass cleanup region, we are able to transfer the $X(\nu''=1)$ population to $X(\nu''=0)$ with {$>90\%$} efficiency. The resulting data is plotted in Fig. \ref{fig:scatt_rate} (blue squares). In this case the decay time is much longer,
since it takes 33 photons for a $1/e$ decay in ground state population when only the $A(\nu'=0)\leftarrow X(\nu''=0)$ is addressed, while it takes $\sim400$ photons when the repump is added. However, the precision of this experiment is limited by the measured VBR values from Sec. \ref{sec:VBR}. From the decay constant of the exponential fit, we obtain a maximum scattering rate {$1.6(1.2)\times10^6$~s$^{-1}$}. The uncertainty mostly comes from the
VBR value $q_{02}$. Nevertheless, the two independent measurements provide an order-of-magnitude estimate of the scattering rate. The relatively high values of $R_{\mathrm{sc}}$ indicate that we can achieve sufficiently high scattering rates for CaH molecules. 
Finally, at the longest interaction time, we estimate that {$170_{-70}^{+500}$} photons per molecule are scattered. 

\section{Magnetically Assisted Sisyphus Cooling}
\label{sec:sisyphus}

\begin{figure*}[ht!]
\begin{centering}
\includegraphics[scale = 1]{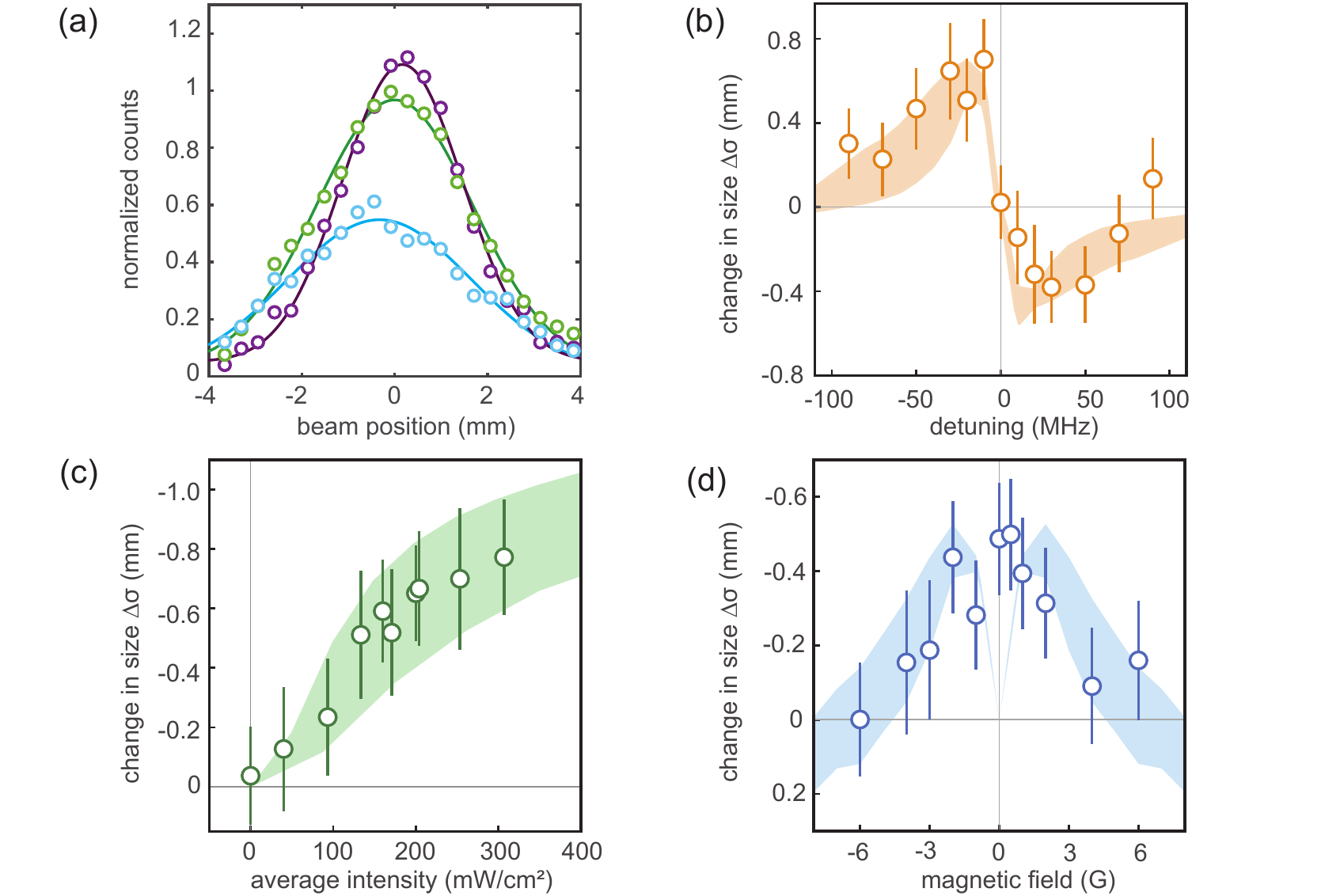}
\caption{Magnetically assisted Sisyphus effect in CaH. (a) Molecular beam profiles obtained for an unperturbed beam (green), Sisyphus cooled beam (purple) at {$\delta = $ +20~MHz}, and Sisyphus heated beam (blue) at {$\delta = $ -20~MHz}. The $y$-axis is normalized to the unperturbed beam maximum and no other scale factors are used. Lines are fits to a 1D Gaussian function to obtain the $1/e$ cloud radius $\sigma$. The increase in on-axis molecule number seen in the cooling configuration is a clear signature of Sisyphus cooling. (b) Change in $\sigma$ as a function of detuning, where $\Delta\sigma>0$ implies heating and vice versa. The detuning is applied globally to each spin-rotation and hyperfine addressing lasers. The data was taken at an intensity of {200~mW/cm$^2$}. The band represents the result of OBE and MC simulations for our experimental system. (c) Change in beam size $\Delta\sigma$ as a function of laser intensity. Detuning has been fixed at $\delta$ = +20~MHz. We do not saturate the Sisyphus cooling effect even at the largest available laser intensity {$(\sim300$~mW/cm$^2)$}. The simulations shown as a band suggest that an intensity of {$>$600~mW/cm$^2$} is required for saturation. (d) Change in beam size $\Delta\sigma$ as a function of magnetic field strength. Detuning is $\delta$ = +20~MHz and intensity is {200~mW/cm$^2$}. Maximum cooling is seen for $B\approx1$~G. Note that the scattering rate is maximized at $B\approx3$~G. The Sisyphus effect is expected to be nulled at $B=0$, but due to the presence of the Earth's magnetic field and the low laser intensity we do not resolve the dip. The simulation shows the same behavior. The bands shown in simulations encompass the spatial variation in laser intensity we expect in the experiment. Each point is a result of 200 repetitions of the experiment, and the experimental error bars are standard errors of Gaussian fitting.}
\label{fig:sisyphus}
\end{centering}
\end{figure*}
The techniques of radiative slowing and magneto-optical trapping rely on the Doppler mechanism, where the scattering rate is optimized when the laser detuning matches the Doppler shift of the molecular transition $(\delta=\vec{k}\cdot\vec{v})$. However, the process of Doppler cooling is fundamentally limited by the excited state lifetime, leading to a minimum achievable temperature at the Doppler limit, $T_D = \hbar/(2k_B\tau)$. For CaH cooled on the $A\leftarrow X$ transition, we estimate $T_D = 116~\mu$K. Hence in order to achieve deep laser cooling of CaH, sub-Doppler cooling techniques must be implemented \cite{Truppe_DopplerLimitMolCooled_2017,Anderegg_2018_grey_molasses_CaF,Caldwell_2019_deep_cooling,Ding_2020_SubDoppler_YO}. Here we demonstrate the ability to perform a type of sub-Doppler cooling known as magnetically assisted Sisyphus cooling in one dimension.

The technique of Sisyphus cooling was first demonstrated with atoms \cite{Emile_1993_sisyphus,Sheehy_1990_sisyphus}. It was subsequently demonstrated with diatomic  \cite{Shuman_LaserCoolingDiatomic_2010,Lim_CooledYbF_2018}, triatomic \cite{Kozyryev_SisyphusSrOH_2017}, and symmetric top \cite{Mitra_CaOCH3Sisphus_2020} molecules. Briefly, molecules travel at a velocity $v$ through a standing wave formed by counter-propagating, near-resonant laser beams. When  the laser is blue-detuned, molecules lose energy as they travel up a Stark potential hill. At the top of the hill where the intensity is highest, molecules absorb the near-resonant photons and rapidly
decay to a dark state, finding themselves at the bottom of the hill. If the magnetic field induced remixing rate is matched to the propagation time along the standing wave, $\lambda/4v$, the molecules return to the bright state and can climb up the potential hill again. This process repeats multiple times, leading to cooling. The opposite effect of Sisyphus heating can be generated by using a red-detuned laser.    

We perform Sisyphus cooling and heating by allowing the laser beam in a multi-pass configuration to overlap between adjacent passes. In order to achieve higher intensities, we keep the laser beam waist relatively small. This leads to substantial beam expansion as the beam propagates. We rely on this expansion after {$\sim16$} passes to create sufficient overlap for a standing wave. We estimate a peak intensity of {$\sim200$~mW/cm$^2$} for one beam within a {5~cm} long interaction region (see Appendix \ref{subsubsec:intensity}). We apply a magnetic field $\vec{B}$ perpendicular to both the molecular beam and the laser wave vector $\vec{k}$, and tune the linear laser polarization to maximize $R_{\mathrm{sc}}$. When optimized, we observe Sisyphus cooling at a detuning of {+20~MHz} as a visible compression of the width of the molecular distribution and also a slight enhancement in the on-axis molecule number (Fig. \ref{fig:sisyphus}(a)). When the detuning is switched to {-20~MHz}, we instead see an increase in the molecular width and the emergence of bimodality near the center, a tell-tale sign of Sisyphus heating. We fit each trace to a 1D Gaussian function to obtain the $1/e$ cloud radius $\sigma$ (see Appendix \ref{sec:OBE}).  

We perform optical Bloch equation (OBE) simulations of the internal states of the molecule in order to estimate the Sisyphus force. Details of the simulation can be found in Refs. \cite{Devlin_2016_OBE,Devlin_2018_OBE} and in Appendix \ref{sec:OBE}. Briefly, we account for
12 ground states and 4 excited states.
We let these molecular states evolve under the OBEs. The force is calculated once the excited state population has reached steady state. Next, we perform Monte Carlo (MC) simulations of individual trajectories as the molecules travel through the interaction region and arrive in the detection region. The spatial distribution from the MC simulation can be compared to the measured camera images. In addition, the associated velocity distribution gives us access to the beam transverse temperature. Furthermore, we consider the full possible range of the standing wave intensity which determines the magnitude of the Sisyphus effect, and use this range to estimate the simulation uncertainty.

We characterize the Sisyphus effect in our experiment as a function of three parameters:  detuning $\delta$, intensity $I$, and magnetic field strength $B$ (Figs. \ref{fig:sisyphus}(b-d)). To quantify the cooling effect, we plot the change in cloud radius, $\Delta\sigma$, measured in mm. To minimize systematic effects, we take one molecule image with the Sisyphus laser beams on in one ablation pulse, followed by one molecule image with them off in the subsequent ablation pulse. This allows us to account for drifts in the ablation yield and beam velocity. We repeat this process for 200 shots to obtain the signal-to-noise ratio depicted in Fig. \ref{fig:sisyphus}. We observe the expected Sisyphus behavior with detuning that is opposite of the Doppler effect:  red-detuned heating and blue-detuned cooling. We additionally observe that the Sisyphus effect persists for detunings up to $\pm50$~MHz  (Fig. \ref{fig:sisyphus}(b)). We next measure the dependence on the laser intensity by varying the laser power while keeping the detuning fixed at $\delta=+20$~MHz. We note that we do not reach saturation of Sisyphus cooling at our maximum available laser intensity. From the simulations, we predict that saturation can be expected for intensities above {600~mW$/$cm$^2$} (Fig. \ref{fig:sisyphus}(c)). At the intensity where we see the largest cooling effect, we estimate that the transverse temperature of the molecular beam is reduced from {12.2(1.2)~mK to 5.7(1.1)~mK} while scattering {$140^{+400}_{-60}$} photons. Lastly, we measure the dependence on magnetic field strength at a fixed detuning ($\delta$ =+20~MHz) and intensity ({200~mW$/$cm$^2$}). The magnetically assisted Sisyphus effect should operate at non-zero magnetic fields, and at our low laser intensities the peak is expected at $\sim1$~G as corroborated by simulations (Fig. \ref{fig:sisyphus}(d)). Since the Earth's field is not cancelled in the experiment, we do not detect a clear dip around $B=0$. Nevertheless, we can be certain that Sisyphus cooling is observed here, since maximum photon scattering occurs at $B\sim3$~G.

\section{Conclusion}
\label{Sec:Conclusion}

In conclusion, we have characterized the dynamics of a cryogenic beam of CaH and experimentally measured the vibrational branching ratios to the first three vibrational levels. We estimate that repumping the $\nu''=1$ and 2 vibrational states should allow us to scatter the $\sim2\times10^4$ photons needed to slow the molecular beam to within the MOT capture velocity. We have demonstrated an ability to scatter {$\sim200$} photos at a rate of {$\sim1.6\times10^6$}~photons$/$s on the $A\leftarrow X$ transition while repumping the first excited vibrational state through the $B\leftarrow X(v''=1)$ transition. This scattering rate implies that, with an additional $\nu''=2$ repumping laser, we should be capable of slowing the molecular beam to within the MOT capture range in {$\sim20$~ms}. Finally we have demonstrated a sub-Doppler cooling mechanism on a CaH beam, reducing the transverse temperature from {12.2(1.2)~mK} to {5.7(1.1)~mK} while only scattering {140} photons via the magnetically assisted Sisyphus effect. Thus we have established that CaH molecules are amenable to further laser cooling. Once these molecules are cooled and trapped in a MOT, they could be used as a precursor for producing dilute ultracold hydrogen via photodissociation, for high-precision fundamental measurements.

\section*{Acknowledgements}
We thank B. Iritani, E. Tiberi, and K. H. Leung for providing a stable laser referenced to the ${^1S}_0 \rightarrow {^3P}_1$ transition in Sr, which was used for wavemeter calibration. We thank K. Wenz, O. Grasdijk, and C. Hallas for fruitful discussions on OBE simulations. This work was supported by the W. M. Keck Foundation grant CU19-2058, ONR grant N00014-21-1-2644, and AFOSR MURI grant FA9550-21-1-0069.

\bibliographystyle{apsrev}
\bibliographystyle{unsrt}


\section{Appendices}
\subsection{Laser configuration}\label{sec:lasers}
For the $A\leftarrow X(v''=0)$ transition at 695~nm, we use two home-built external cavity diode lasers (ECDLs) separated by $\sim$ 2~GHz to address the $J=1/2$ and $J=3/2$ manifolds. Each ECDL is then passed through an acousto-optic modulator to generate two frequencies separated by the hyperfine splitting of the corresponding $J$-manifold (54~MHz for $J=1/2$ and 101~MHz for $J=3/2$, Fig. \ref{fig:Mol_Structure}(d)). The resulting four frequencies are used to individually seed four injection-locked amplifiers (ILAs). Laser beams from the ILAs corresponding to a single $J$-manifold are first combined with orthogonal linear polarizations on a polarizing beam splitter, and then the two $J$-manifolds are combined on a 50:50 beam splitter. Hence a single $\lambda/2$-waveplate is sufficient to determine the polarization of each frequency component. The combined beam is spatially overlapped with light addressing the $B\leftarrow X(v''=1)$ repump transition at 690~nm using a narrow-band dichroic filter (FF01-690/8-25). Each repump transition is addressed with light produced by two ECDLs, each seeding one ILA and addressing a $J$-manifold. The hyperfine sidebands are added to the seed light via electro-optic modulators (EOMs), resonant at $\sim$ 50~MHz and using different order Bessel functions. In total, this laser setup is capable of providing 150~mW of cycling light and 110~mW of repumping light propagating from the same fiber. To achieve the maximum intensity for Sisyphus experiments, an additional 150~mW of cooling light was added to the system though a separate fiber. 
After cooling we repump any left over $v''=1$ population to $v''=0$ using 40~mW of repump light in the cleanup region. Detection is performed on the $B\leftarrow X(v''=0)$ transition by using two ECDLs at 635~nm addressing the two $J$-manifolds, with the hyperfine sidebands added via EOMs. Since 60~mW of light is sufficient for detection, no ILAs are used. 

\subsection{VBR measurement}\label{sec:filters}

Here we present the details that factor into the calculation of $A$ and $B$ state VBRs (Sec. \ref{sec:VBR}). In general, $R_{\nu''}$, where $\nu''=0,1,2$, is the fitted ratio of two PMT time traces. The fitting time window is from 1~ms to 7~ms, and we use the data between 35~ms and 90~ms for background subtraction. 
We use the same fitting protocol for all $R_{\nu''}$ measurements. The ratio $R_{\nu''}$ is stable during data collection and only varies if the position of either PMT is altered.

The quantum efficiencies of PMTs are either experimentally measured or obtained from factory calibration results. We use the following expression to measure the quantum efficiency:
\begin{equation}
    Q = \frac{\text{Cts}}{P/\hbar\omega},
\end{equation}

\noindent where Cts is the total number of PMT counts per second, $P$ is the laser power incident into the PMT, and $\hbar\omega$ is the photon energy. Each PMT is placed into a black box, and an optical fiber carrying light directly points at the PMT active surface. To prevent saturation of the PMT, we insert calibrated neutral-density filters between the fiber and the PMT head. We measure $Q$ at several different laser powers and fit to a line to obtain the PMT linear response.  
Eventually we measured $Q$ at 635~nm, 690~nm and 695~nm, and find that our measured $Q$ is within $5\%$ of the manufacturer's specifications. However, due to the lack of available laser sources at the other fluorescence wavelengths, we employ the factory calibrated values for $Q$ provided by the manufacturer, and assign a $5\%$ error to them. We also directly measure the transmission efficiencies of all dichroic filters used for the experiment if a laser source is available, otherwise we use the manufacturer's specifications.

Table \ref{tab:VBR_table_appendix} shows all the measured values that are used in calculating VBRs and their errors, and Table \ref{table:dichroic_table_appendix} lists the dichroic filters used for the measurement of VBRs.

\begin{table}[htbp]
\begin{tabular}{ccclccc}
\cline{1-3} \cline{5-7}
\multicolumn{1}{|c|}{\textbf{$A^2\Pi_{1/2}~~q_{01}$}} &
  \multicolumn{1}{c|}{\textbf{value}} &
  \multicolumn{1}{c|}{\textbf{error}} &
  \multicolumn{1}{l|}{} &
  \multicolumn{1}{c|}{\textbf{$A^2\Pi_{1/2}~~q_{02}$}} &
  \multicolumn{1}{c|}{\textbf{value}} &
  \multicolumn{1}{c|}{\textbf{error}} \\ \cline{1-3} \cline{5-7} 
\multicolumn{1}{|c|}{$R_0$} &
  \multicolumn{1}{c|}{1.575} &
  \multicolumn{1}{c|}{0.013} &
  \multicolumn{1}{l|}{} &
  \multicolumn{1}{c|}{$R_0$} &
  \multicolumn{1}{c|}{1.363} &
  \multicolumn{1}{c|}{0.015} \\ \cline{1-3} \cline{5-7} 
\multicolumn{1}{|c|}{$R_1$} &
  \multicolumn{1}{c|}{0.0410} &
  \multicolumn{1}{c|}{0.0017} &
  \multicolumn{1}{l|}{} &
  \multicolumn{1}{c|}{$R_2$} &
  \multicolumn{1}{c|}{0.0007} &
  \multicolumn{1}{c|}{0.0005} \\ \cline{1-3} \cline{5-7} 
\multicolumn{1}{|c|}{$Q_{P_2, \lambda_{01}} /  Q_{P_2, \lambda_{00}}$} &
  \multicolumn{1}{c|}{0.73} &
  \multicolumn{1}{c|}{0.04} &
  \multicolumn{1}{l|}{} &
  \multicolumn{1}{c|}{$Q_{P_2, \lambda_{02}} /  Q_{P_2, \lambda_{00}}$} &
  \multicolumn{1}{c|}{0.167} &
  \multicolumn{1}{c|}{0.008} \\ \cline{1-3} \cline{5-7} 
\multicolumn{1}{|c|}{$T_{F_3,\lambda_{01}}/ T_{F_2, \lambda_{00}}$} &
  \multicolumn{1}{c|}{1.17} &
  \multicolumn{1}{c|}{0.06} &
  \multicolumn{1}{l|}{} &
  \multicolumn{1}{c|}{$T_{F_3,\lambda_{02}}/ T_{F_2, \lambda_{00}}$} &
  \multicolumn{1}{c|}{1.15} &
  \multicolumn{1}{c|}{0.06} \\ \cline{1-3} \cline{5-7} 
\multicolumn{1}{|c|}{\textbf{$q_{01}/q_{00}$}} &
  \multicolumn{1}{c|}{\textbf{0.0306}} &
  \multicolumn{1}{c|}{\textbf{0.0025}} &
  \multicolumn{1}{l|}{} &
  \multicolumn{1}{c|}{\textbf{$q_{02}/q_{00}$}} &
  \multicolumn{1}{c|}{\textbf{0.0025}} &
  \multicolumn{1}{c|}{\textbf{0.0019}} \\ \cline{1-3} \cline{5-7} 
\multicolumn{1}{l}{} &
  \multicolumn{1}{l}{} &
  \multicolumn{1}{l}{} &
   &
  \multicolumn{1}{l}{} &
  \multicolumn{1}{l}{} &
  \multicolumn{1}{l}{} \\ \cline{1-3} \cline{5-7} 
\multicolumn{1}{|c|}{\textbf{$B^2\Sigma^+~~q_{01}$}} &
  \multicolumn{1}{c|}{\textbf{value}} &
  \multicolumn{1}{c|}{\textbf{error}} &
  \multicolumn{1}{l|}{} &
  \multicolumn{1}{c|}{\textbf{$B^2\Sigma^+~~q_{02}$}} &
  \multicolumn{1}{c|}{\textbf{value}} &
  \multicolumn{1}{c|}{\textbf{error}} \\ \cline{1-3} \cline{5-7} 
\multicolumn{1}{|c|}{$R_0$} &
  \multicolumn{1}{c|}{5.83} &
  \multicolumn{1}{c|}{0.05} &
  \multicolumn{1}{l|}{} &
  \multicolumn{1}{c|}{$R_0$} &
  \multicolumn{1}{c|}{5.83} &
  \multicolumn{1}{c|}{0.03} \\ \cline{1-3} \cline{5-7} 
\multicolumn{1}{|c|}{$R_1$} &
  \multicolumn{1}{c|}{0.0696} &
  \multicolumn{1}{c|}{0.0021} &
  \multicolumn{1}{l|}{} &
  \multicolumn{1}{c|}{$R_2$} &
  \multicolumn{1}{c|}{0.0040} &
  \multicolumn{1}{c|}{0.0005} \\ \cline{1-3} \cline{5-7} 
\multicolumn{1}{|c|}{$Q_{P_2, \lambda_{01}} /  Q_{P_2, \lambda_{00}}$} &
  \multicolumn{1}{c|}{0.86} &
  \multicolumn{1}{c|}{0.05} &
  \multicolumn{1}{l|}{} &
  \multicolumn{1}{c|}{$Q_{P_2, \lambda_{02}} /  Q_{P_2, \lambda_{00}}$} &
  \multicolumn{1}{c|}{0.56} &
  \multicolumn{1}{c|}{0.03} \\ \cline{1-3} \cline{5-7} 
\multicolumn{1}{|c|}{$T_{F_3,\lambda_{01}}/ T_{F_2, \lambda_{00}}$} &
  \multicolumn{1}{c|}{1.01} &
  \multicolumn{1}{c|}{0.04} &
  \multicolumn{1}{l|}{} &
  \multicolumn{1}{c|}{$T_{F_3,\lambda_{02}}/ T_{F_2, \lambda_{00}}$} &
  \multicolumn{1}{c|}{1.00} &
  \multicolumn{1}{c|}{0.05} \\ \cline{1-3} \cline{5-7} 
\multicolumn{1}{|c|}{\textbf{$q_{01}/q_{00}$}} &
  \multicolumn{1}{c|}{\textbf{0.0137}} &
  \multicolumn{1}{c|}{\textbf{0.0011}} &
  \multicolumn{1}{l|}{} &
  \multicolumn{1}{c|}{\textbf{$q_{02}/q_{00}$}} &
  \multicolumn{1}{c|}{\textbf{0.00125}} &
  \multicolumn{1}{c|}{\textbf{0.00019}} \\ \cline{1-3} \cline{5-7} 
\end{tabular}
\caption{Measured parameter values that are used to calculate VBRs and FCFs for CaH given in Sec. \ref{sec:VBR}.}
\label{tab:VBR_table_appendix}
\end{table}

\begin{table}[htbp]
\begin{tabular}{c|c|c|c|}
\cline{2-4}
                                                   & \textbf{$F_1$} & \textbf{$F_2$} & \textbf{$F_3$} \\ \hline
\multicolumn{1}{|c|}{\textbf{$A^2\Pi_{1/2}~~q_{01}$}}    & FF01-692/40-25       & FF02-684/24-25       & FF01-760/12-25       \\ \hline
\multicolumn{1}{|c|}{\textbf{$A^2\Pi_{1/2}~~q_{02}$}}    & FF01-692/40-25       & FF02-684/24-25       & FF01-840/12-25       \\ \hline
\multicolumn{1}{|c|}{\textbf{$B^2\Sigma^+~~q_{01}$}} & FL635-10       & FF01-630/20-25       & FF01-690/8-25        \\ \hline
\multicolumn{1}{|c|}{\textbf{$B^2\Sigma^+~~q_{02}$}} & FL635-10       & FF01-630/20-25       & FF01-760/12-25       \\ \hline
\end{tabular}
\caption{List of all dichroic filters used in the VBR and FCF measurements for CaH given in Sec. \ref{sec:VBR}.}
\label{table:dichroic_table_appendix}
\end{table}

\subsection{OBE and MC simulations}
\label{sec:OBE}

We developed the optical Bloch equation solver \cite{Devlin_2016_OBE, Devlin_2018_OBE, KonradWenz_2021_thesis, KonradWenz_2020_BCF} using Python and Julia (via PyJulia). The source code can be found online\footnote{github.com/QiSun97/OBE-Solver}. We include 12 ground states of $|X^2\Sigma^+,\ \nu''=0,\ N''=1\rangle$ in Hund's case (b) and 4 excited states of $|A^2\Pi_{1/2},\ \nu'=0,\ J'=1/2\rangle$ in Hund's case (a). We ignore another 12 states in the $|X^2\Sigma^+,\ \nu''=1,\ N''=1\rangle$ level, because the population in the vibrationally excited state is not significant in our experiment. The transition dipole moments are calculated with the help of a Matlab package\footnote{github.com/QiSun97/Rabi$\_$Matrix$\_$Elements$\_$Calculator} where a Hund's case (b) basis is projected onto a case (a) basis. We perform Monte-Carlo simulation of the classical trajectories of the cryogenic molecular beam\footnote{github.com/QiSun97/Cryogenic$\_$Beam$\_$Sisyphus$\_$MC$\_$Simulation}. We initialize $10^4$ molecules at the exit of the 5~mm beam aperture described in Sec. \ref{sec:setup}, and propagate them through the interaction region where they experience Sisyphus forces as described in Sec. \ref{sec:sisyphus}. 

We combine the OBE and MC simulations as follows: at a given laser polarization and other experimental parameters, we perform a two-dimensional parameter sweep of velocity and laser intensity using the OBE simulation of the optical force. 
Then within the MC simulation, for each particle at given spatial position and velocity, we perform a 2D interpolation to obtain the instantaneous force on the particle.
In order to obtain an accurate spatial variation of intensity, we consider Gaussian beam expansion and power loss per pass due to imperfections. We measure the beam width before it enters the interaction region, and use the number of passes to estimate the traveling distance and calculate the laser beam waist. The beam waist $w(z)$ at a distance $z$ is calculated as
\begin{equation}\label{eq:waist}
    w(z) = w(0)\sqrt{1 + (z/z_R)^2}
\end{equation}
where $z_R = \pi w(0)^2 n /\lambda$ is the Rayleigh range. We measure the spacing of the laser beams to convert from the spatial coordinate to the number of passes. We also observe a moderate power loss every time a laser beam passes through the chamber ($\sim1.8\%$). Together, these provide us with a conversion from spatial position to local laser intensity. Molecules propagate through the interaction region and eventually exit to reach the detection region. We then plot the spatial distribution of molecules and perform a one-dimensional Gaussian fit to extract the width information. The fit function used for all experimental data as well as simulations is
\begin{equation}
    y(r) = y(0) + A\exp{\left(-(r-r_0)^2/2\sigma^2\right)}
\end{equation}
\noindent where $\sigma$ is the $1/e$ radius of the cloud (Fig. \ref{fig:sisyphus}(a)).

\subsubsection{Assignment of simulation uncertainty}
\label{subsubsec:band}

The main source of uncertainty in the simulations stems from our inability to measure the position and amplitude of the standing wave that gives rise to the Sisyphus effect. Although the multi-pass laser beams are distinguishable initially, after 16 passes they overlap significantly. This overlap region, where the standing wave is formed and Sisyphus forces act, covers $\sim5$~cm of the interaction length. In addition, it is challenging to estimate the beam waist within this region. We quantify this uncertainty by considering two situations:  (1) the laser beams are tightly spaced and the effective overlap is long, and (2) the beams are loosely spaced and their overlap is small. For example, in Fig. \ref{fig:sisyphus}(c), the simulation band ranges from a 5~cm overlap with 1~mm beam spacing, or a 4~cm overlap with 2~mm beam spacing. The same strategy is used in generating the simulation bands in Figs. \ref{fig:sisyphus}(b,d) as well.    

\subsubsection{Laser intensity estimation} \label{subsubsec:intensity}

The intensity of a Gaussian laser beam is defined as $I = 2P/\pi w^2$ where $I$ is the peak intensity, $P$ is the total power, and $w$ is the $1/e^2$ waist. We measure the laser beam power before it enters the interaction region, and then estimate the peak intensity after the beam undergoes $N$ passes using Eq. (\ref{eq:waist}). Furthermore, since the beam overlap that can lead to Sisyphus effect is between the beam aperture and $\sim$ 5~cm downstream, we denote the average intensity between 7~cm and 12~cm from the first pass of the laser beam that is coupled from the downstream side of the 12~cm long interaction region (Fig. \ref{fig:Mol_Structure}(a)). This is how the $x$-axis of the experimental data in Fig. \ref{fig:sisyphus}(c) is generated.

For the MC simulation, we define the average intensity from the local intensities experienced by each molecule as the molecular beam traverses the interaction region. We tabulate the local intensities experienced by all detected particles at the end of the simulation and calculate the median of the distribution to obtain the average intensity.

\subsubsection{Transverse temperature estimation}

We estimate the transverse temperature of the molecular beam as follows. Within the MC simulation, the only free parameter that allows us to match the detected spatial distribution is the transverse temperature that governs the transverse velocity distribution. The spatial distribution is assumed to be uniform at the 5~mm aperture, and the forward velocity is experimentally determined. Thus, we obtain a one-to-one correspondence of the molecular beam width to the transverse temperature of the beam. The unperturbed beam has a width of 3.11(14)~mm, which corresponds to 12.2(1.2)~mK, and the coldest beam has a width of 2.34(13)~mm, which corresponds to 5.7(1.1)~mK.

\end{document}